\documentclass{article}
\usepackage{spconf,amsmath,graphicx}
\usepackage{booktabs}
\usepackage{color,amsmath,url,times, tabularx,bbm,amssymb,multirow}

\title{A Mutual learning framework for Few-shot Sound Event Detection}

\name{Dongchao Yang$^{1}$,
      Helin Wang$^{1}$,
      Yuexian Zou$^{1,*}$\thanks{This paper was partially supported by the Shenzhen Science \& Technology Fundamental Research Programs (No: JCYJ20180507182908274 \& JSGG20191129105421211) and GXWD20201231165807007-20200814115301001.}\thanks{$^{*}$ Corresponding Author: zouyx@pku.edu.cn},
      Zhongjie Ye$^{1}$, 
      Wenwu Wang$^2$}
\address{$^{1}$ADSPLAB, School of ECE, Peking University, Shenzhen, China\\
$^2$Center for Vision, Speech and Signal Processing, University of Surrey, UK
}
\begin{document}
%
\maketitle
\begin{abstract}
Although prototypical network (ProtoNet) has proved to be an effective method for few-shot sound event detection, two problems still exist. Firstly, the small-scaled support set is insufficient so that the class prototypes may not represent the class center accurately. Secondly, the feature extractor is task-agnostic (or class-agnostic): the feature extractor is trained with base-class data and directly applied to unseen-class data. 
To address these issues, we present a novel mutual learning framework with transductive learning, which aims at iteratively updating the class prototypes and feature extractor. More specifically, we propose to update class prototypes with transductive inference to make the class prototypes as close to the true class center as possible. To make the feature extractor to be task-specific, we propose to use the updated class prototypes to fine-tune the feature extractor. After that, a fine-tuned feature extractor further helps produce better class prototypes.
Our method achieves the F-score of 38.4$\%$ on the DCASE 2021 Task 5 evaluation set, which won the first place in the few-shot bioacoustic event detection task of Detection and Classification of Acoustic Scenes and Events (DCASE) 2021 Challenge.
\end{abstract}
\section{Introduction}
\label{sec:intro}
Deep learning-based sound event detection methods typically require large amounts of data for training or fine-tuning models for specific applications \cite{parascandolo2016recurrent,kong2020panns,dinkel2021towards}. The development of deep learning models to detect unseen sound classes with only few labels is insufficient. Recently, studies \cite{pons2019training,chou2019learning} have proposed to tackle this problem using few-shot learning (FSL), where a classifier needs to learn to recognize novel classes given only few samples of each class. In the FSL setting, a model is first trained on labeled data with base classes. Then, model generalization is evaluated on few-shot tasks, composed of unlabeled samples from novel classes unseen during training (query set), assuming only one or a few labeled samples (support set) are given per novel class. Prototypical network (ProtoNet) \cite{snell2017prototypical} has been proved as an effective method for few-shot sound event detection \cite{wang2020few,shi2020few}. In DCASE 2021 Challenge Task 5, the official baseline and several solutions \cite{anderson2021bioacoustic,tangtwo} submited to this challenge have also employed ProtoNet. However, there are still two factors that limit the performance of ProtoNet. Firstly, the class feature of the support set may be insufficient due to the presence of background noise and interference in audio data, so that the class prototypes learned from such support set may not represent the class center accurately. Figure 1 shows the learned representations (embeddings) extracted from ProtoNet, and we can see that the embeddings of each class are scattered, especially for the support set of `buk4.wav', which contains more background noise than `a1.wav'. As a result, the F-score of `buk4.wav' is much lower than that of `a1.wav'. Secondly, ProtoNet trains a feature extractor with the base-class data and applies the feature extractor to samples from unseen classes. This style of transfer learning is task-agnostic: the feature extractor is not learned to be optimally discriminative with respect to the unseen classes. It often performs worse than a task-specific feature extractor \cite{ye2020few,bateni2020improved}.
\begin{figure}[t]
  \centering
  \includegraphics[width=\linewidth]{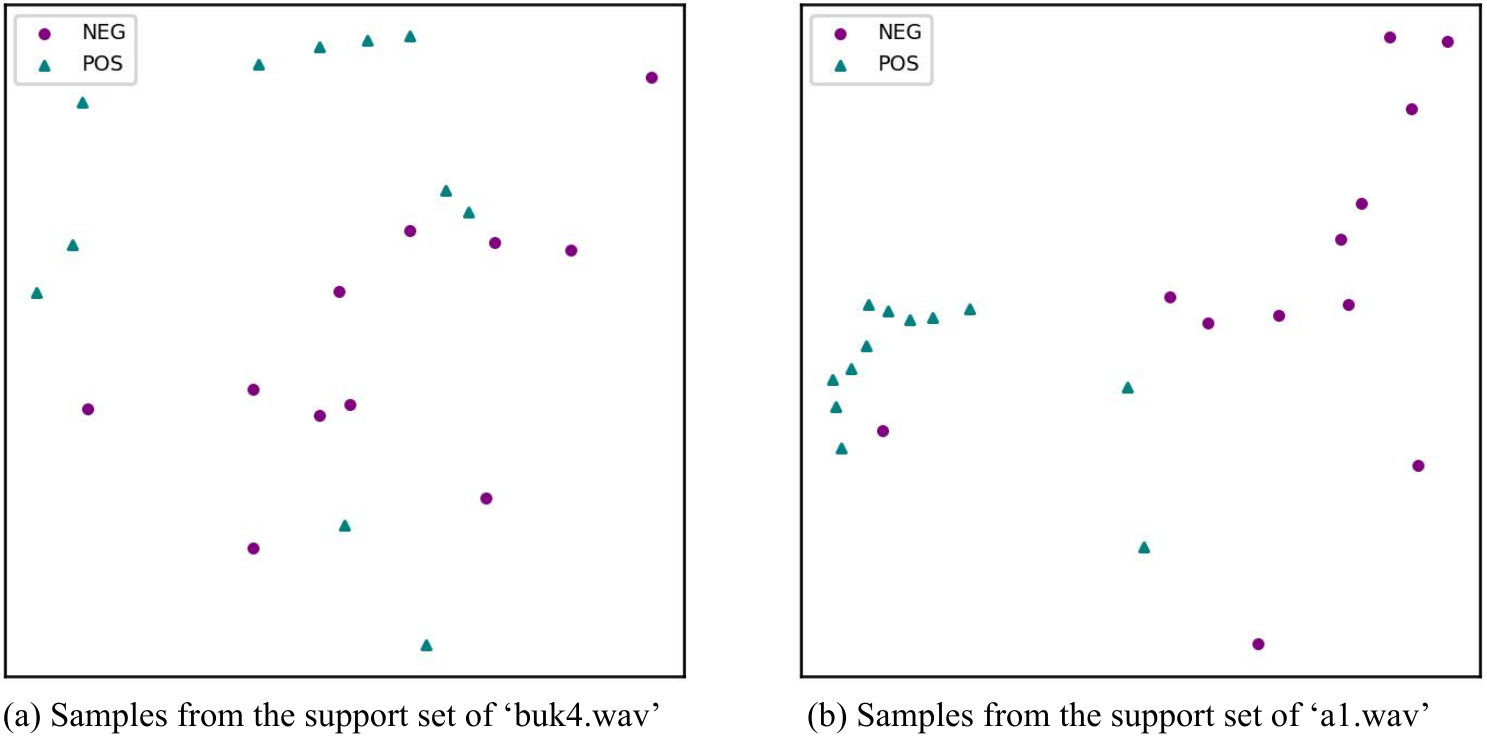}
  \caption{The visualization of the embeddings of the support set from test audio file by t-SNE \cite{van2008visualizing}. We choose two audios (`buk4.wav' and `a1.wav'), each audio includes two classes (POS and NEG). The overall F-score values of `buk4.wav' and `a1.wav' are 45.5\% and 63.7\% respectively.}
  \label{fig:distribution}
\end{figure}
Following the discussions and observations above, we propose a mutual learning framework to continuously update the feature extractor and class prototypes.
More specifically, we firstly train a feature extractor with base-class data and use the class prototypes to initialize a classifier.
We then leverage the statistics of unlabelled audio to update the classifier with transductive inference \cite{boudiaf2020transductive,boudiaf2020few,liu2018learning}.
In order to obtain a task-specific feature extractor, we further use the updated class prototypes as the supervised information to fine-tune the feature extractor.
These processes can be repeated several times so that the feature extractor and classifier can be continuously updated.
Our contributions can be summarized as follows: (1) To solve the problem that class prototypes cannot represent the true class centers accurately, we propose to update class prototypes with transductive learning. (2) To make the feature extractor to be task-specific, we propose a novel method to fine-tune the feature extractor. (3) Our mutual learning framework significantly improves the performance of few-shot bioacoustic event detection over the state-of-the-art methods.
\section{Proposed method}
\label{sec:proposed method}
In this section, few-shot setting, transductive inference and the mutual learning framework will be introduced.
\subsection{Few-shot setting} 
Assume we are given a training set, 
$X_{\mathit{base}}=\left\{ \boldsymbol{x}_i,\boldsymbol{y}_i \right\}^{N_{\mathit{base}}}_{i=1}$, where $\boldsymbol{x}_i$ denotes the acoustic feature of example $i$ and $\boldsymbol{y}_i$ denotes associated one-hot label. Let $Y_{\mathit{base}}$ denote the label set of this base dataset. The few-shot learning assumes that we are given a test dataset: $X_{\mathit{test}}=\left\{ \boldsymbol{x}_i,\boldsymbol{y}_i \right\}^{N_{\mathit{test}}}_{i=1}$, with a completely new label set $Y_{test}$ such that $Y_{\mathit{base}} \cap Y_{\mathit{test}}=\emptyset$, and the test dataset only has a few labelled examples. 
\subsection{Transductive inference}
Transductive inference (TI) is about reasoning from observed, specific (training) cases to specific (test) cases. In this paper, the core idea of TI is about leveraging the statistics of the unlabeled data. More specifically, we adapt the idea from \cite{boudiaf2020transductive}, which maximizes the mutual information (MI) between the query features and their label predictions for a few-shot task at inference. It means that the model has seen these unlabeled data before making final prediction.
\begin{figure}[t]
  \centering
  \includegraphics[width=\linewidth]{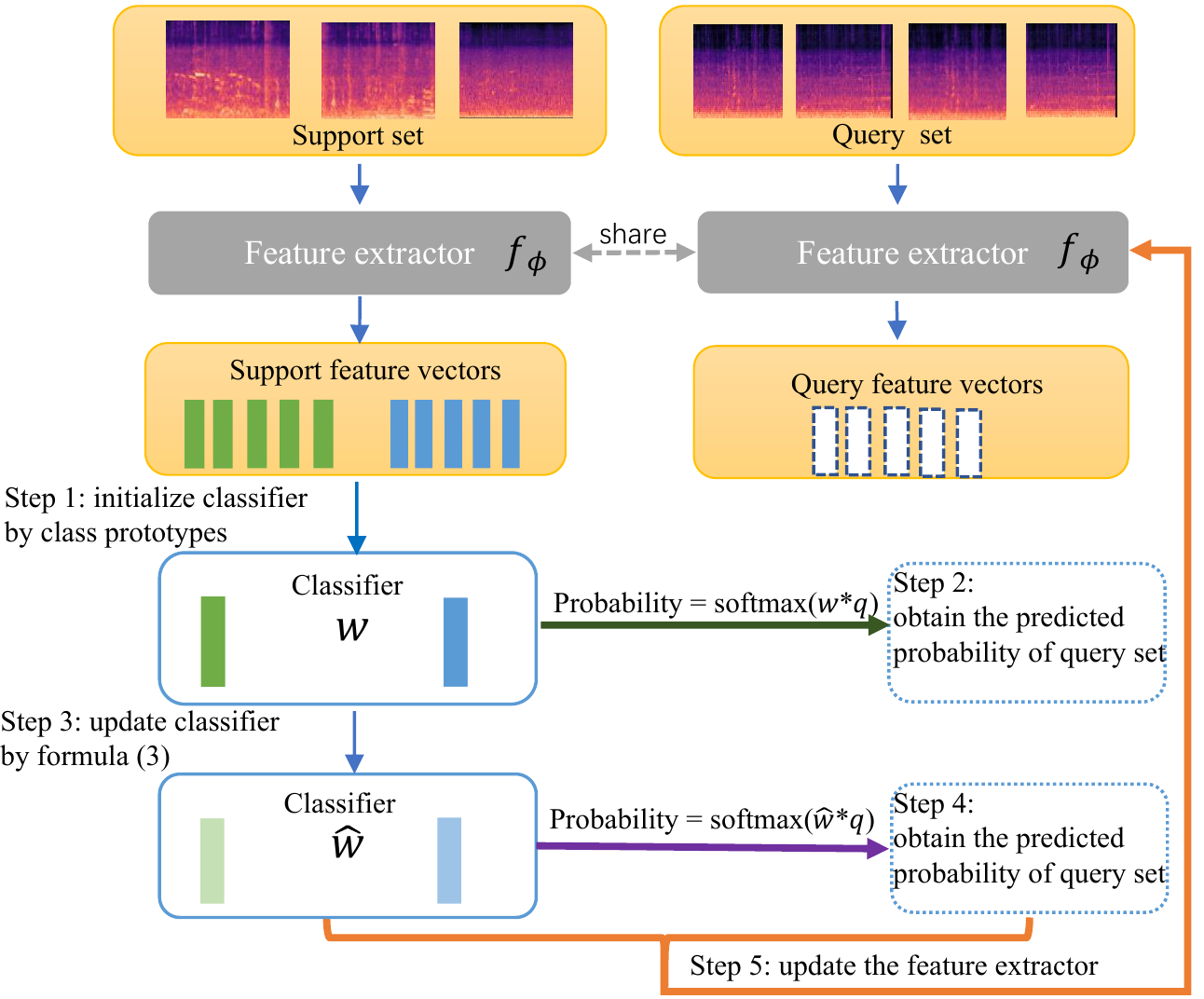}
  \caption{The overview of mutual learning framework. Feature extractor is trained with base class data.}
  \label{fig:ml}
\end{figure}
\subsection{Mutual learning framework}
The overview of the mutual learning framework is shown in Figure 2. In this section, we first introduce how to use class prototypes to build a classifier and update the classifier with transductive inference. After that, we discuss how to make use of the updated class prototypes to fine-tune the feature extractor. Lastly, we summarize the core idea of mutual learning framework.\\
\textbf{Building classifier} For a given few-shot task, with a support set $S$ and a query set $Q$, let $X$ denote the random variables associated with the acoustic features within $S \cup Q$ and let $Y=\left\{1,2, ..., K \right\}$ be the random variables associated with the labels. Let $f_{\phi} : X \rightarrow Z \subset R^d$ denote the encoder (\textit{i.e.}, feature extractor) function of a deep neural network, where $\phi$ denotes the trainable parameters, and $Z$ stands for the set of embedded features. The encoder is firstly trained from the base training set $X_{\mathit{base}}$ using the standard cross-entropy loss. Next, for each specific few-shot task, 
we define a classifier, parametrized by a weight matrix $\boldsymbol{W} =[\boldsymbol{w}_1,...,\boldsymbol{w}_K] \in R^{K \times d} $. The posterior distribution over labels given features is defined by $p_{ik}=\mathbb{P}(Y=k|X=\boldsymbol{x}_i;\boldsymbol{W},\phi)$. The marginal distribution over query labels is defined by $\hat{p}_k=\mathbb{P}(Y_Q=k;\boldsymbol{W},\phi)$. $p_{ik}$ and $\hat{p}_k$ are calculated as formula (\ref{formula calculate p}).
\begin{equation}\label{formula calculate p}
\setlength{\abovedisplayskip}{4pt}
\setlength{\belowdisplayskip}{4pt}
        p_{ik} = \frac{\mathit{exp}(\boldsymbol{w}_k \cdot \boldsymbol{z}_i)}{\sum_{c=1}^{K}\mathit{exp}(\boldsymbol{w}_{c} \cdot \boldsymbol{z}_i)},
    \hat{p}_k=\frac{1}{Q} \sum_{i \in Q}p_{ik}
\end{equation}
where $\boldsymbol{z}_i = \frac{f_{\phi}(\boldsymbol{x}_i)}{||f_{\phi}(\boldsymbol{x}_i)||_2}$ denotes L2-normalized embedded features. For each task, weights $\boldsymbol{W}$ are initialized by the class prototypes of the support set, as follows
\begin{equation}\label{w init}
\setlength{\abovedisplayskip}{4pt}
\setlength{\belowdisplayskip}{4pt}
\begin{split}
   \boldsymbol{w}_k = \frac{1}{|S_k|} \sum_{\boldsymbol{x}_i \in S_k} f_{\phi}(\boldsymbol{x}_i) 
   \end{split}
\end{equation}
In this paper, we only need to judge whether the audio frame is a positive sample, so we set $K=2$.\\
\textbf{Updating classifier} To update the weight matrix $\boldsymbol{W}$, for each single few-shot task,
we propose a loss function with two complementary terms: (1) a standard cross-entropy loss on the support set; (2) a mutual-information loss, which includes a conditional entropy loss and a marginal entropy loss.
\begin{equation}\label{formula feed-forward2}
\begin{split}
   L_{w} = \lambda_{\mathit{CE}} \cdot \mathit{CE} - \hat{I}(X_Q;Y_Q)  \\
   \end{split}
\end{equation}
\begin{equation}\label{formula feed-forward}
\begin{split}
  \mathit{CE} = -\frac{1}{|S|}\sum_{i \in S}\sum_{i=1}^K y_{ik}log(p_{ik}) \\
   \end{split}
\end{equation}
\begin{equation}\label{formula MI}
\begin{split}
    \hat{I}(X_Q;Y_Q) = -\sum_{k=1}^K \hat{p}_k log\hat{p}_k +  \frac{1}{|Q|}\sum_{i \in Q}\sum_{k=1}^K p_{ik}log(p_{ik})
   \end{split}
\end{equation}
where $\mathit{CE}$ denotes the cross entropy loss function, $y_{ik}$ denotes the true label of the sample in the support set, $p_{ik}$ denotes the prediction result. In our experiments, $\lambda_{\mathit{CE}}$ is set as 0.1. $\hat{I}(X_Q;Y_Q)$ denotes the mutual information between the query samples and their latent labels. It is a combination of two terms, the first term is the empirical label-marginal entropy, denoted as $\hat{H}(Y_Q)$, while the second term is an empirical estimate of the conditional entropy of labels given the query acoustic features, denoted as $\hat{H}(Y_Q|X_Q)$. $\hat{H}(Y_Q|X_Q)$ aims at minimizing the uncertainty of the posteriors at unlabelled query samples, thereby encouraging the model to output confident predictions. This entropy loss is widely used in the context of semi-supervised learning (SSL) \cite{grandvalet2005semi,berthelot2019mixmatch}, as it models effectively the cluster assumption: the classifier’s boundaries should not occur at dense regions of the unlabelled features. The label-marginal entropy regularizer $\hat{H}(Y_Q)$ encourages the marginal distribution of labels to be uniform.\\
Note that we only update the weight matrix $\boldsymbol{W}$ in this step, while the feature extractor is fixed. Our experimental results also show that simultaneously updating feature extractor $f_{\phi}$ and weight matrix $\boldsymbol{W}$ does not offer better performance.\\
\noindent
\textbf{Updating feature extractor} Previous works \cite{ye2020few,bateni2020improved} have shown that a task-specific feature extractor works better than a task-agnostic one, so we expect our feature extractor to be task-specific.
To achieve this, we propose a novel method to update feature extractor,
which uses the updated class prototypes as supervision information to fine-tune the feature extractor. 
In addition, we plan to make use of the predicted results for unlabelled data. Figure \ref{fig:ml} shows the updating process of our method. 
When we finish step 3 and 4, we make use of $\hat{\boldsymbol{W}}$ and predicted results of high confidence to fine-tune the feature extractor $f_{\phi}$. 
The loss function has two terms as formula (\ref{update f}) shows, including a cross-entropy (CE) loss according to pseudo label and a contrastive loss.
\begin{equation}\label{update f}
\setlength{\abovedisplayskip}{4pt}
\setlength{\belowdisplayskip}{4pt}
\begin{split}
   L_f = \lambda_{1} * \mathit{CE} + \lambda_{2} * L_{c}
   \end{split}
\end{equation}
where $\lambda_{1}$ and $\lambda_{2}$ are hyper-parameters. In our experiments, $\lambda_{1}=\lambda_{2}=0.5$. The contrastive loss $L_c$ is defined as follows.
\begin{equation}\label{contrast loss}
\begin{split}
   L_c = -log(\frac{\mathit{exp}(\mathit{sim}(\hat{\boldsymbol{w}}[1],\boldsymbol{\Bar{z}}_\mathit{pos}))}{\sum_{k=1}^{N} \mathit{exp}(\mathit{sim}(\hat{\boldsymbol{w}}[1],z_{\mathit{neg}}^{k}))})
\end{split}
\end{equation}
where $\hat{\boldsymbol{w}}[1]$ denotes the first row vector of $\hat{\boldsymbol{W}}$, and it represents the prototype of positive class in our experiments. $\boldsymbol{\Bar{z}}_\mathit{pos}$ denotes the mean of the learned representation of the positive samples on the support set, and $z_{\mathit{neg}}$ denotes the learned representation of negative samples. $N$ denotes the number of negative samples. In our experiments, $\mathit{sim}$ stands for cosine similarity. We do not use $\hat{\boldsymbol{w}}[2]$ for the reason that the negative sample is randomly chosen. This loss function also can be viewed as knowledge distillation \cite{hinton2015distilling}.\\
\noindent
\textbf{Mutual learning} According to previous discussion, we can make use of transductive inference to improve the classifier, and we can also improve feature extractor by the updated classifier and pseudo label. After we get a better feature extractor, we can continue running the previous process to update the classifier. It means that feature extractor and classifier can learn from each other, so we name it as mutual learning.

\section{Experiment}
\label{sec:experiments}
\subsection{Experimental setups}
\textbf{Dataset} The dataset is from DCASE2021 task 5 \cite{morfi_veronica_2021_4543504}, including development and evaluation sets. The development set is pre-splitted into training and validation sets. The training set contains about 14 hours of audio, and the validation set contains 5 hours of audio. The evaluation set consists of 31 audio files acquired from different bioacoustic sources.

\noindent
\textbf{Metrics} For all the experiments, we use the event-based F-measure \cite{mesaros2016metrics} as the evaluation metric, which is one of the most commonly used metrics for sound event detection.

\noindent
\textbf{Preprocessing} All the raw audios are down-sampled to 22.05 kHz and applied a Short Time Fourier Transform (STFT) with a window size of 1024 samples, followed by a Mel-scaled filter bank on perceptually weighted spectrograms. This results in 128 Mel frequency bins and around 86 frames per second. The input frames are normalized to zero-mean and unit variance according to the training set.

\noindent
\textbf{Training} We use the same backbone as the baseline \cite{snell2017prototypical}, which only includes 4 convolutional layers. The only difference is that we do not use meta-learning training strategy. Instead, we directly train feature extractor by the cross entropy loss. Specifically, we use a dense layer after the feature extractor, and then add a softmax layer to get classification probability. The Adam optimizer is applied for a total of 15 epochs, with an initial learning rate of $1 \times 10^{-3}$.

\noindent
\textbf{Updating classifier} The test audio only gives the first five positive annotations, and negative samples are randomly sampled from unlabelled parts. In order to update the classifier $\boldsymbol{W}$, the Adam optimizer is used for a range of 5-30 epochs, with an initial learning rate of $1 \times 10^{-5}$. We choose different training epoch for different test audio, for the reason that training epochs will affect the prediction results. The prediction results at the last epoch are used as our final results.
\begin{center}
\begin{table}[t] \centering
\caption{F-score comparison of different methods on DCASE 2021 task5 Development and Evaluation dataset.}
\label{tab:my-table-eval}
\begin{tabular}{|c|c|c|}
\hline
Method          & Dev-set        & Eval-set      \\ \hline
Baseline \cite{snell2017prototypical}        & 41.48          & 20.1          \\ \hline
Anderson \textit{et al.} \cite{anderson2021bioacoustic} & 26.2  & 35.0 \\ \hline
Tang \textit{et al.} \cite{tangtwo} & 51.4  & 38.3 \\ \hline
\textbf{TI (ours)}    & 51.21          & 33.2          \\ \hline
\textbf{TI-ML (ours)} & \textbf{55.26} & \textbf{38.4} \\ \hline
\end{tabular}
\end{table}
\end{center}
\begin{table}[t] \centering
\caption{Ablation study on the effect of each term in formula (\ref{formula feed-forward2}). CE: Cross entropy loss, I: Mutual information loss.}
\label{tab:table2}
\begin{tabular}{cllll}
\hline
\multicolumn{1}{l}{Method}                & Loss    & Precision & Recall & F-score \\ \hline
\multicolumn{1}{c|}{\multirow{3}{*}{TI}} & \textit{None}      & 16.89     & 60.1  & 26.38   \\
\multicolumn{1}{c|}{} & I      & 57.8     & 43.9  & 49.96   \\
\multicolumn{1}{c|}{}                     & CE    & 55.17     & 46.07  & 50.21   \\
\multicolumn{1}{c|}{}                     & I+CE    & 57.11     & 46.39  & \textbf{51.21}  \\ \hline
\multicolumn{1}{c|}{\multirow{3}{*}{TI-ML}} & \textit{None}      & 15.5     & 50.5  & 23.76   \\
\multicolumn{1}{c|}{} & I      & 69.6     & 43.6  & 53.67   \\
\multicolumn{1}{c|}{}                     & CE    & 52.68     & 49.1  & 50.83   \\
\multicolumn{1}{c|}{}                     & I+CE    & 65.54     & 47.76  & \textbf{55.26}  \\ \hline
\end{tabular}
\end{table}
\vspace{-1cm}

\noindent
\textbf{Updating feature extractor} We build a new dense layer after the feature extractor, which only need to do binary classification task. The Adam optimizer is used for a total of 5 epochs, with an initial learning rate of $1 \times 10^{-4}$ for feature extractor, and $1 \times 10^{-3}$ for the new dense layer.
\subsection{Experimental results}
Table \ref{tab:my-table-eval} shows the experimental results. Our method achieves 38.4 $\%$ F-score on evaluation set, which significantly outperforms the baseline \cite{snell2017prototypical}. TI denotes we only use transductive learning to update classifier (class prototypes). TI-ML denotes we use mutual learning framework to update classifier and feature extractor. TI-ML performs better than TI, which shows the effectiveness of our mutual learning framework. Anderson \textit{et al.} \cite{anderson2021bioacoustic} also applied ProtoNet, and compared with baseline \cite{snell2017prototypical}, their model utilized both per-channel energy normalisation (PCEN) on the front end and three data augmentation methods. In contrast, our approach does not need to employ these strategies and gets a better result. In addition, Tang \textit{et al.} \cite{tangtwo} got a very close result with us on the evaluation dataset, but they used a 12-layer ResNet pre-trained on AudioSet \cite{gemmeke2017audio} as the backbone.
\begin{table}[t] \centering
\caption{Ablation study on the effect of iterations on mutual learning.}
\label{tab:my-table}
\begin{tabular}{ccccc}
\hline
Method                                       & Iterations & Precision     & Recall         & F-score        \\ \hline
\multicolumn{1}{c|}{\multirow{4}{*}{TI-ML}} & 0         & 57.12         & 46.39          & 51.21          \\
\multicolumn{1}{c|}{}                        & 1         & 65.54         & 47.76 & \textbf{55.26} \\
\multicolumn{1}{c|}{}                        & 2         & 72.50 & 43.38          & 54.28          \\
\multicolumn{1}{c|}{}                        & 3         & 69.53         & 43.68          & 53.66          \\ \hline
\end{tabular}
\end{table}

\subsection{Ablation study}
In this part, we discuss the influence of transductive learning and mutual learning. The experiments were carried out on the development set.\\
\textbf{Influence of each term on formula (\ref{formula feed-forward2})} We now assess the impact of each term in formula (\ref{formula feed-forward2}). The results are reported in Table 2. We observe that integrating the two terms in our loss consistently outperforms any other configuration. \textit{None} indicates that we do not update classifier $\boldsymbol{W}$, otherwise we directly use the class prototypes to initialize the classifier, and then use it to predict results. We first analyze their effect on transductive inference (TI). If we do not update the classifier, the F-score is 26.38\%. The performance is lower than baseline \cite{snell2017prototypical} for the reason that we do not use the meta-learning training strategy. On the contrary, when we use either cross entropy loss or mutual information loss to update the classifier, the performance will be significantly improved. We can find that only using cross-entropy loss brings improvement. It proves that insufficient support set leads to the problem that class prototypes cannot represent the true class center, because class prototypes cannot even properly classify the support set. Secondly, we analyze their effect on the mutual learning (TI-ML). If we use the classifier which is not updated to fine-tune feature extractor, we will get the worst results. If we only use the cross entropy loss to update the classifier, it brings a small improvement. Compared with the cross entropy loss, the mutual information loss has more advantages on mutual learning. The best results can be obtained when both two terms are used.  \\
\textbf{Influence of each term on formula (6)} To fine-tune the feature extractor, we devise a loss function as shown in (6) which includes a cross the entropy loss and a contrastive loss. If we only use the cross entropy loss, the result is 51.64\%. If the contrastive loss is used alone, the result is 51.06\%. The best result (55.26\%) can be obtained when both are used, which shows that these two loss terms are both important.\\
\textbf{Influence of iterations} Table 3 reports the impact of iterations for mutual learning. Here, the iteration equals to 0 means we do not update the feature extractor. Experimental results indicate updating it only once offers the best F-score. We find that the performance tends to decline when the number of iterations increase, one of the reasons is that the number of negative samples is far more than that of positive samples in this dataset, which makes the model learn more information about negative samples, so the result shows higher False Positives (FP) rate.
\section{Conclusions}
\label{sec:conclusion}
In this paper, we proposed a mutual learning framework with transductive inference to continuously improve the ability of feature extractor and classifier.
Our method won the first place in the DCASE 2021 Challenge Task 5, with a F-score of 38.4$\%$.
In the future, we will further improve the performance of our systems. 
The source code is released.\footnote{https://github.com/yangdongchao/DCASE2021Task5}



\bibliographystyle{IEEE.bst}
\bibliography{refs.bib}

\end{document}